\documentclass[fleqn,usenatbib]{mnras}

\usepackage{newtxtext,newtxmath}
\usepackage[T1]{fontenc}
\usepackage{graphicx}	% Including figure files
\usepackage{amsmath}	% Advanced maths commands
\usepackage[version=4]{mhchem}
\usepackage{lipsum}
\usepackage[english]{babel}
\usepackage[utf8]{inputenc}
\usepackage{soul}
\usepackage{booktabs}
\usepackage{multirow}
\usepackage{hyperref}
\usepackage{color}
\newcommand*\xbar[1]{%
   \hbox{%
     \vbox{%
       \hrule height 0.5pt % The actual bar
       \kern0.5ex%         % Distance between bar and symbol
       \hbox{%
         \kern-0.1em%      % Shortening on the left side
         \ensuremath{#1}%
         \kern-0.1em%      % Shortening on the right side
       }%
     }%
   }%
} 

\title[Stability of Multiply Charged Naphthalene]{Multiply charged naphthalene and its \ce{C10H8} isomers: bonding, spectroscopy and implications in AGN environments}

 \author[Santos et al.]{Julia C. Santos,$^{1,2}$\thanks{E-mail: santos@strw.leidenuniv.nl (J.C.S.)}$\ddagger$ Felipe Fantuzzi,$^{3}$\thanks{E-mail: f.fantuzzi@kent.ac.uk (F.F.)}\thanks{Contributed equally.} Heidy M. Quitián-Lara,$^{3,4}$ Yanna Martins-Franco,$^{4}$  
\newauthor
Karín Menéndez-Delmestre,$^{4}$ Heloisa M. Boechat-Roberty,$^{4}$ and Ricardo R. Oliveira$^1$\thanks{E-mail: rrodrigues.iq@gmail.com (R.R.O.)}
\\
$^{1}$Instituto de Qu\'{i}mica, Universidade Federal do Rio de Janeiro, Av. Athos da Silveira Ramos 149, 21941909 Rio de Janeiro, Brazil. \\
$^{2}$Laboratory for Astrophysics, Leiden Observatory, Leiden University, PO Box 9513, NL–2300 RA Leiden, The Netherlands. \\
$^{3}$School of Physical Sciences, Ingram Building, University of Kent, Park Wood Rd, Canterbury CT2 7NH, United Kingdom. \\
$^{4}$Observatório do Valongo, Universidade Federal do Rio de Janeiro, Ladeira do Pedro Antônio 43, 20080-090 Rio de Janeiro, Brazil. \\
}

\date{Accepted XXX. Received YYY; in original form ZZZ}

\pubyear{2021}

\hypersetup{draft}
\begin{document}
\label{firstpage}
\pagerange{\pageref{firstpage}--\pageref{lastpage}}
\maketitle

\begin{abstract}

\noindent Naphthalene (C$_{10}$H$_{8}$) is the simplest polycyclic aromatic hydrocarbon (PAH) and an important component in a series of astrochemical reactions involving hydrocarbons. Its molecular charge state affects the stability of its isomeric structures, which is specially relevant in ionised astrophysical environments. We thus perform an extensive computational search for low-energy molecular structures of neutral, singly, and multiply charged naphthalene and its isomers with charge states +$q$ = 0$-$4 and investigate their geometric properties and bonding situations. We find that isomerisation reactions should be frequent for higher charged states and that open chains dominate their low-energy structures. We compute both the scaled-harmonic and anharmonic infrared spectra of selected low-energy species and provide the calculated scaling factors for the naphthalene neutral, cation, and dication global minima. All simulated spectra reproduce satisfactorily the experimental data and, thus, are adequate for aiding observations. Moreover, the potential presence of these species in the emission spectra of the circumnuclear regions of active galactic nuclei (AGNs), with high energetic X-ray photon fluxes, is explored using the experimental value of the naphthalene photodissociation cross-section, $\sigma_{ph-d}$, to determine its half-life, $t_{1/2}$, at a photon energy of 2.5 keV in a set of relevant sources. Finally, we show that the computed IR bands of the triply and quadruply charged species are able to reproduce some features of the selected AGN sources.

\end{abstract}


%%%%%%%%%%%%%%%%%%%%%%%%%%%%%%%%%%%%%%%%%%%%%%%%%%
%%%%%%%%%%%%%%%%% BODY OF PAPER %%%%%%%%%%%%%%%%%%

\section{Introduction}
\label{intro}

The recent detections of highly polar aromatic molecules, namely cyanocyclopentadiene ($c$-C$_5$H$_5$CN, \citealt{McCarthy2021a,KelvinLee2021}), benzonitrile ($c$-C$_6$H$_5$CN, \citealt{McGuire2018}), and cyanonaphthalene (CNN, \citealt{McGuire2021,McCarthy2021a}), in Taurus Molecular Cloud 1 (TMC-1) have been encouraging the search for other aromatic derivatives in interstellar and circumstellar environments. Indeed, in addition to these nitrogen-bearing species, pure hydrocarbon cycles, such as cyclopentadiene (\textit{c}-C$_5$H$_6$, \citealt{Cernicharo2021a}), indene (\textit{c}-C$_9$H$_8$, \citealt{Burkhardt2021a,Cernicharo2021a}), and \textit{ortho}-benzyne (\textit{o}-C$_6$H$_4$, \citealt{Cernicharo2021b}), were also recently and unambiguously detected in TMC-1. Benzyne and benzonitrile are proxies for benzene \citep{Lee2019,Cooke2020} and can serve as molecular probes. Analogously, cyanonaphthalene can act as a proxy for naphthalene (\ce{C10H8}), a small aromatic molecule featuring fused bicyclic six-membered rings and the simplest polycyclic aromatic hydrocarbon (PAH). The search for naphthalene and its derivatives is not restricted to TMC-1 because it is possible to find aromatic molecules in different astrophysical environments. A very illustrative example is benzonitrile, which was also detected in Serpens 1A, Serpens 1B, Serpens 2, and the dense core MC27/L1521F \citep{Burkhardt2021b}.

Due to the relevance of naphthalene in fields such as astrochemistry, combustion science, and environmental chemistry, several studies have focused on its spectroscopic properties and formation pathways at different charge states. This vast body of knowledge has provided clear evidence that naphthalene is an important intermediate of PAH growth routes \citep{Alliati2019}. Also, through ionisation events, naphthalene can trigger the formation of diacetylene (C$_4$H$_2$) and the benzene radical cation \citep{Solano2015}, thus demonstrating that the charge state influences the formation of other species, including cyclic molecules. Furthermore, even for highly valence-excited states of the naphthalene cation, relaxation pathways to the ground state due to vibronic coupling and conical intersections are present \citep{Marciniak2015,Reitsma2019}. This trend, which is also observed for larger PAHs \citep{Wenzel2020}, corroborates the hypothesis that multiply charged PAHs \citep{tielens08,Zhen2017,Zhen2018,Banhatti2021}, together with their nitrogen-doped, phosphorus-doped, and protonated counterparts \citep{Mattioda2008,Ricks2009,AlvaroGalue2017,Meyer2021,Oliveira2021}, are also viable candidates as carriers of the unidentified infrared (UIR) bands.

In addition to fragmentation processes, the ionisation of small cycles, including aromatic rings, can induce molecular isomerisation \citep{Jasik2015,Monfredini2016,Fantuzzi2018,Quitian-Lara2018,Monfredini2019,Fantuzzi2019,Wolff2020}, which is usually accompanied by a severe loss of molecular integrity. Several examples were demonstrated by screening the potential energy surface (PES) of a given molecular structure for each charge state under consideration \citep{Jasik2015,Fantuzzi2018,Gutsev2020,Hendrix2020,Cerqueira2020}. For benzene, the basic unit of PAHs, the six-membered ring structure is the global minimum for the neutral and singly charged states, but the dicationic singlet ground state has an unusual pentagonal-pyramidal geometry \citep{Krogh-Jespersen1991,Jasik2014,Jasik2015,Fantuzzi2017a,Wolff2020,Pozo-Guerron2020a}. Additionally, while the dicationic global minimum of toluene (C$_7$H$_8$) and aniline (C$_6$H$_7$N) are characterised by the presence of a six-membered ring \citep{Monfredini2016,Forgy2018}, the dicationic chlorobenzene (C$_6$H$_5$Cl) is composed of an unusual structure featuring a cyclopropenyl ring and a terminal C$=$Cl double bond in a formyl-like CHCl moiety \citep{Fantuzzi2018}. Finally, the cationic and dicationic global minima of biphenyl (\ce{C10H12}), another important building block of larger PAHs, are composed of acenaphthene-like structures, where a C$_2$ motif bridges a naphthalene backbone \citep{Quitian-Lara2020}.

Given the prevalence of PAHs at differente charge states in astrophysical environments, and taking into account that the molecular charge state can dramatically influence the relative stability of isomeric structures, herein we investigate geometrical and spectroscopic properties of neutral and multiply charged cations of naphthalene (C$_{10}$H$_8$) and its isomers at different $+q$ charge states ($q=$ 0$-$4). Initially, we perform an extensive search for molecular structures in order to reveal low-energy isomers and geometric patterns of C$_{10}$H$_8$ species for all charge states. Furthermore, we compute scaled-harmonic and anharmonic infrared spectra (IR) of the most representative species, which are discussed in the context of UIR bands. Finally, our findings are also extrapolated to the chemistry of environments with hard radiation fields, where PAHs with a high degree of ionisation have been observed \citep{Witt2006, Peeters2012}, focusing in circumnuclear regions of dusty active galaxy nuclei (AGNs).

The paper is organised as follows: the methodology of geometry and spectroscopic calculations is shown in Section \ref{comp_details}. The results about the multiply charged isomers' stability and their main transitions are described in Section \ref{results_discussion}. In Section \ref{implications} we contextualize our results using mid-IR spectra of active galaxies. Lastly, we sum up our findings in Section \ref{conclusions}.

\section{Computational Methods}
\label{comp_details}

All calculations performed in this work are based on density functional theory (DFT). Initially, we conduct an exhaustive exploration of the chemical space of \ce{C10H8} species at the DFT level for different charge (+$q$) states, i.e., neutral ($q=0$), cation ($q=1$), dication ($q=2$), trication ($q=3$), and tetracation ($q=4$). To speed-up the calculations, we use the resolution-of-the-identity method \citep{Weigend2006} coupled with the chain-of-spheres (COSX) algorithm \citep{Neese2009}. For each charge state, we perform around 185 geometry optimisations and  Hessian evaluations with the PBE0 functional \citep{Adamo1999,Ernzerhof1999} combined with the def2-TZVP basis set \citep{Weigend2005}. The initial structures were generated by the open molecule generator (OMG) software \citep{Peironcely2012}. For all cases the lowest possible multiplicity was chosen (singlet for $q=0,2,4$ and doublet for $q=1,3$). Further single-point calculations were conducted for selected species at higher multiplicities to compute singlet-triplet and doublet-quartet vertical energy gaps. All reported geometries are characterised as minimum energy structures at their corresponding potential energy surfaces by the vibrational frequency analysis as no imaginary frequency was found. All energy values discussed herein are enthalpies computed at the standard temperature of 298.15 K ($H^{298.15}$) and are reported as relative values with respect to the global minimum of each charge state $q$. Similar stability trends are observed for calculations considering only the electronic energy including zero-point energy corrections (see Table S1 in the Supporting Information). The isomers are labelled in the ascending order of $H^{298.15}$, with \textbf{1$^{q+}$} being the corresponding global minimum of the charge state $q$.

In order to analyse the energetic trends of selected isomers as a function of the charge state $q$, we use the deviation plot (D-plot) analysis \citep{Fernandez-Lima2006,Fernandez-Lima2007,Fantuzzi2013}. This analysis consists of grouping species into distinct classes of compounds and calculating their deviation energies $D^{298.15}_{n,q}$ from the average energy ($\bar{H}^{298.15}_{q}$) for each distinct value of $q$:

\begin{equation}
    D^{298.15}_{n,q} = H^{298.15}_{n,q} - \bar{H}^{298.15}_{q}, 
\end{equation}

\noindent where:

\begin{equation}
\bar{H}^{298.15}_{q}  = \sum_{n=1}^{N} \frac{H^{298.15}_{n,q}}{N}.
\end{equation}

The electronic structure and bonding situation of selected species is analysed by inspection of the frontier Kohn-Sham molecular orbitals and additional calculations based on the intrinsic bond orbital (IBO; \citealt{Knizia2013}) and natural bond orbital (NBO; \citealt{Weinhold2016}) methods. In order to investigate the aromaticity (or antiaromaticity) properties of the species, nucleus independent chemical shift (NICS; \citealt{Schleyer1996,Chen2005}) calculations are also performed.

Scaled-harmonic and anharmonic IR spectra of all structures with $H^{298}$ values in-between 0-10 kcal mol$^{-1}$ from the global minima are simulated from further optimisations and frequency calculations at the B3LYP/def2-TZVPP level of theory \citep{Vosko1980,Lee1988,Becke1993,Stephens1994,Weigend2005}. The B3LYP functional was chosen due to its better performance at predicting the frequencies and intensities of the molecular infrared bands in comparison to PBE0. The anharmonic spectra are calculated with the vibrational second-order perturbation (VPT2) method \citep{Nielsen1951,Barone2014}. The NASA Ames PAH IR Spectroscopic Database (PAHdb; \citealt{Bauschlicher2018}) contains the largest library of both laboratory-measured and theoretically-computed PAH spectra. Following its guidelines, we apply three scaling factors to the harmonic spectra, with ranges of $\nu>2500$ cm$^{-1}$ (1), $2500>\nu>1111.\bar{1}$ cm$^{-1}$ (2), and $\nu < 1111.\bar{1}$ cm$^{-1}$ (3). The scaling factors were determined by minimising the following expression:

\begin{equation}
\sum^n_{i=1}(s(i)*\omega_e(i)-\nu(i))^6,
\end{equation}

\noindent where $\omega_e$ is the harmonic frequency, $\nu$ is the experimental fundamental counterpart, $s(i)$ is the scaling factor, and $n$ is the number of IR bands for which there are experimental data. The simulated spectra are convoluted with a Gaussian lineshape function with a FWHM of 15 cm$^{-1}$, in accordance with \cite{Boersma2014}. 

\begin{figure*}
\centering
    \includegraphics[width=\textwidth]{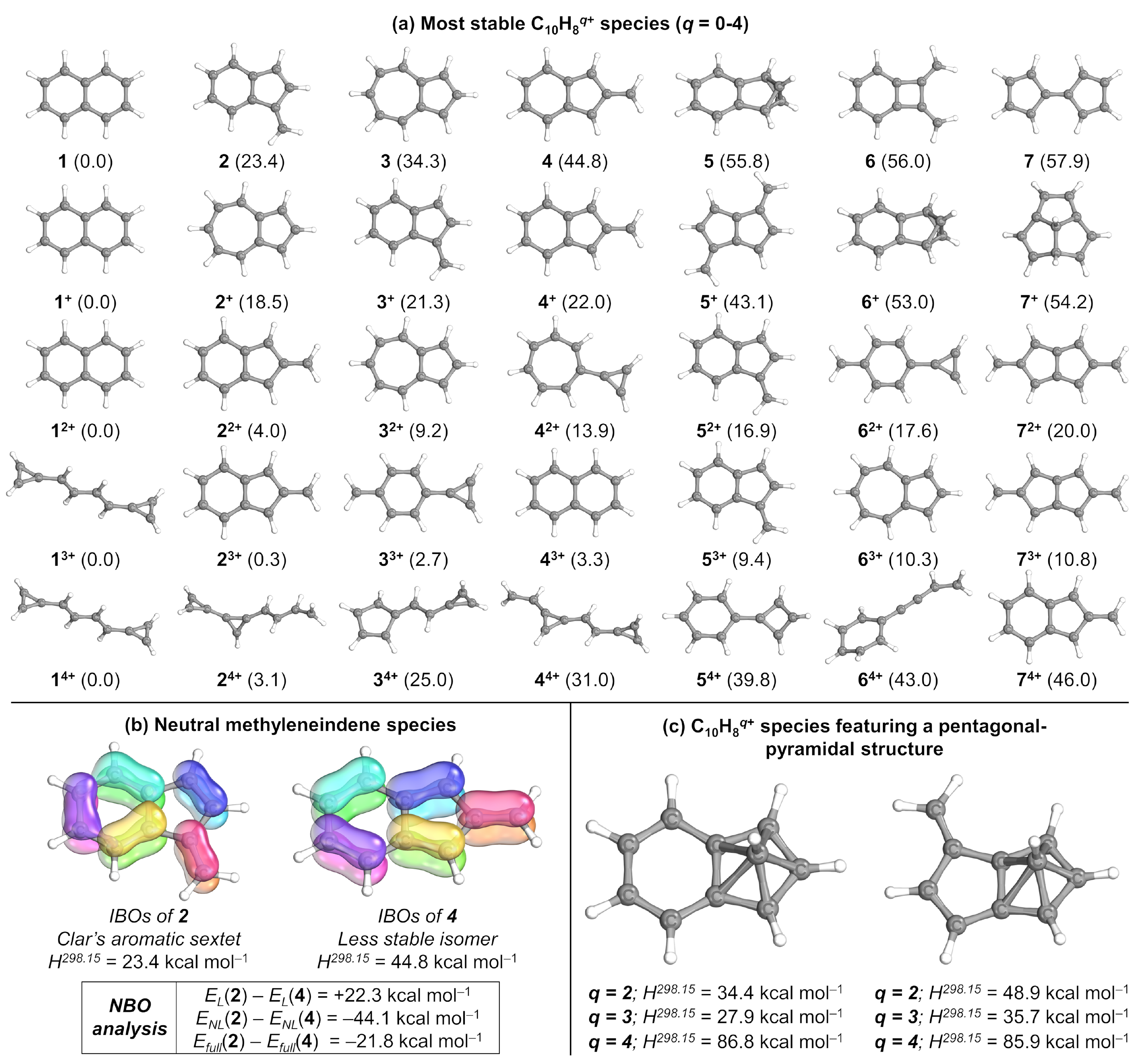} 
    \caption{(a) Low-lying energy \ce{C10H8}$^{q+}$ isomers at neutral, cation, dication, trication and tetracation charge states obtained at the PBE0/def2-TZVP level of theory. Relative enthalpies (kcal mol$^{-1}$) are shown in parenthesis. (b) IBOs, relative energies, and NBO analysis of the neutral methyleneindene species \textbf{2} and \textbf{4}. (c) Structure and energetics of \ce{C10H8}$^{q+}$ species ($q =$ 2$-$4) featuring a pentagonal-pyramidal structure.}
    \label{fig:structures}
\end{figure*}

All geometry optimisations were conducted using Orca 4.2 \citep{Neese2012}. The IBO calculations were done with the IBOview software \citep{Knizia2013}. Scaled-harmonic and anharmonic IR spectra, as well as the NICS calculations, were performed using the Gaussian 16, revision C.01  \citep{g16}.

\section{Results and Discussion}
\label{results_discussion}

\subsection{Bonding and relative stability}

\begin{figure}
\centering
    \includegraphics[width=\columnwidth]{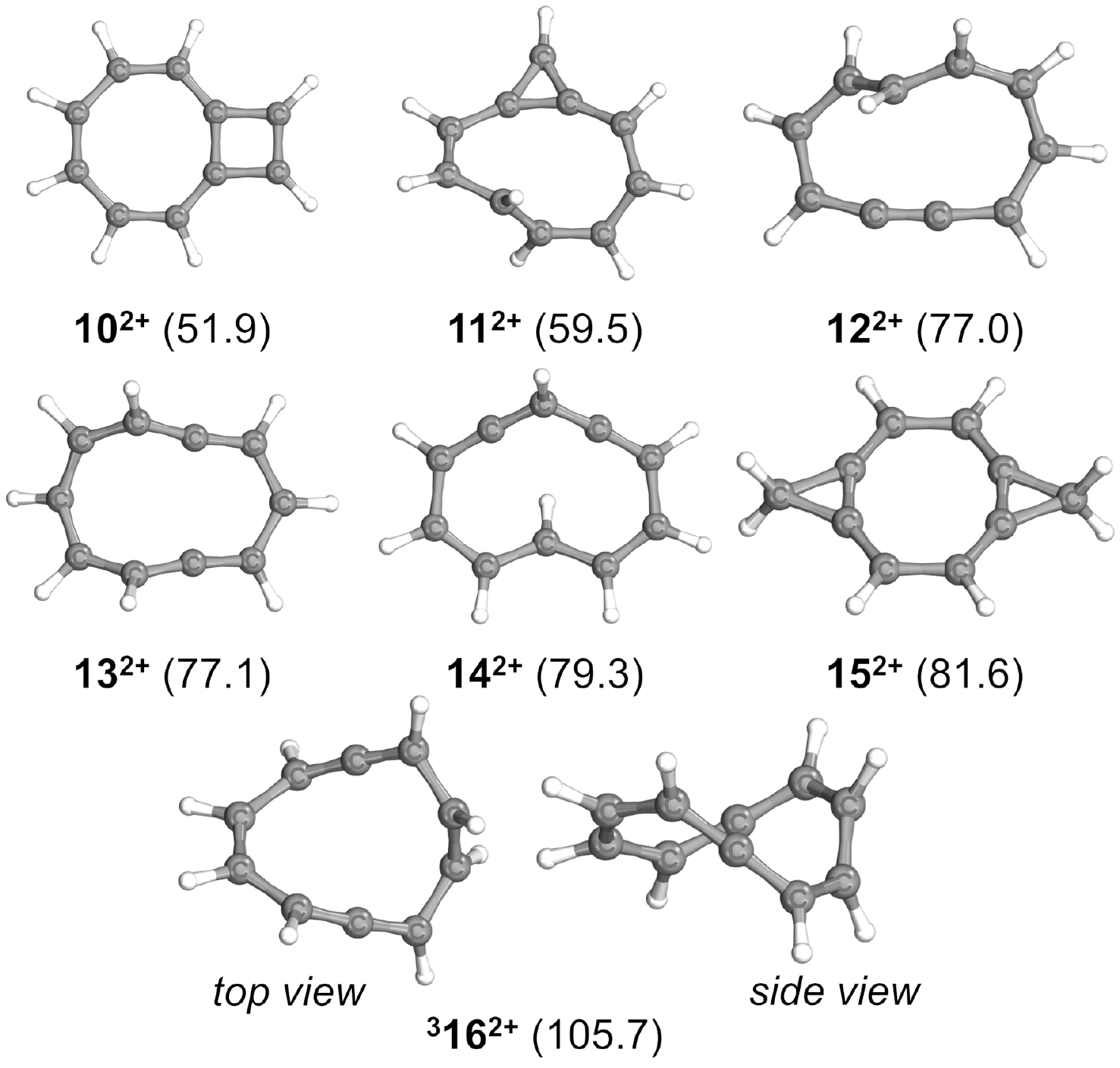} 
    \caption{Selected optimised dicationic structures featuring large rings. Relative enthalpies (kcal mol$^{-1}$) are shown in parenthesis.}
    \label{fig:rings}
\end{figure}

\begin{figure*}
\centering
    \includegraphics[width=\textwidth]{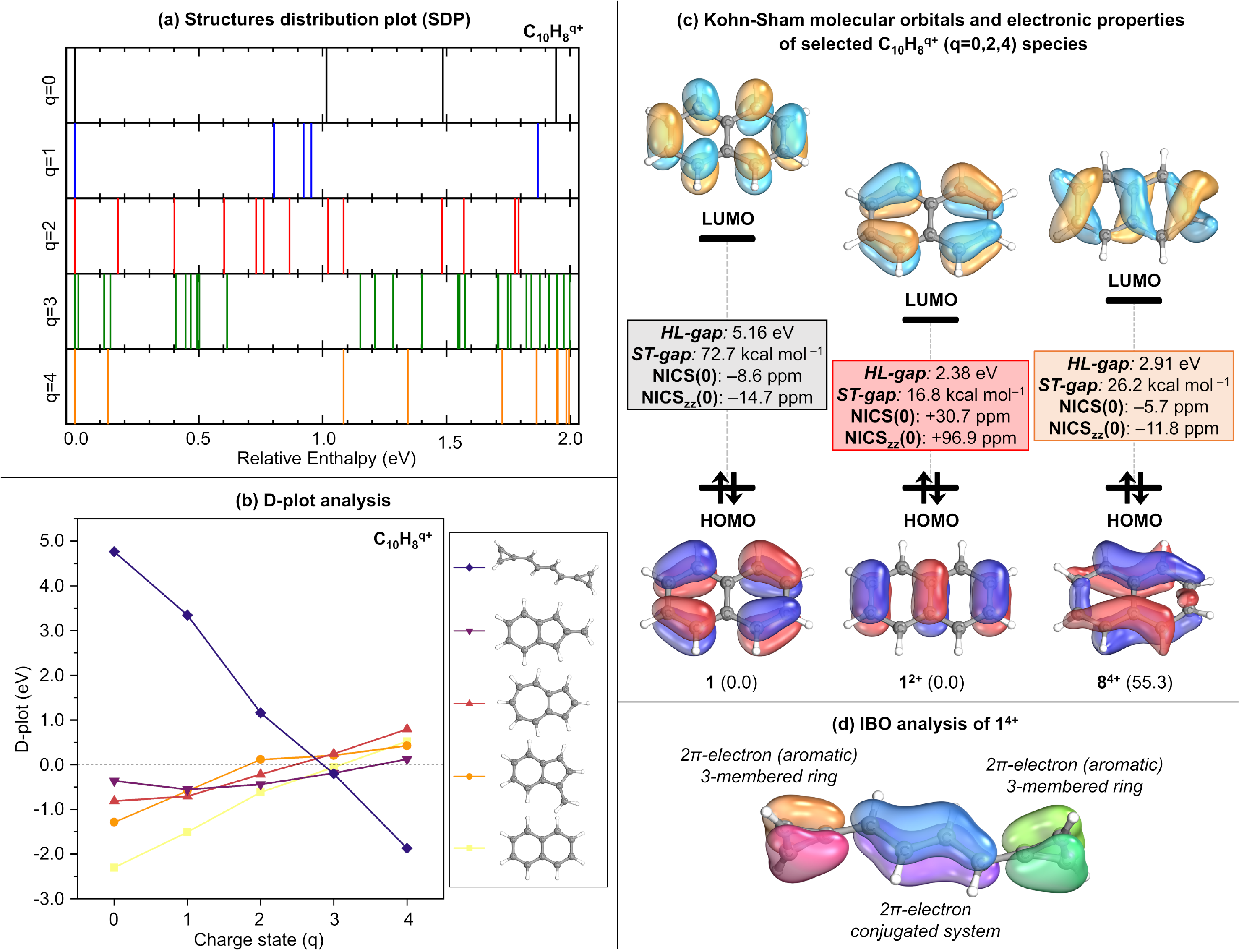} 
    \caption{(a) Structures distribution plot (SDP) of \ce{C10H8}$^{q+}$ isomers ($q=$ 0$-$4) in the energy window of 0 to 2.0 eV from the corresponding global minima. (b) D-plot analysis of selected \ce{C10H8}$^{q+}$ isomers ($q=$ 0$-$4). (c) Canonical Kohn-Sham molecular orbitals of \ce{C10H8}$^{q+}$ isomers ($q=$ 0$-$4) featuring naphthalenic structures. Their corresponding HL-gap, ST-gap, NICS(0), and NICS$_{zz}$(0) are also shown. (d) Bonding situation of the tetracationic \textbf{1$^{4+}$} species from the IBO perspective.}
    \label{fig:bonding}
\end{figure*}

The seven most stable isomers of \ce{C10H8} for each charge state $q = 0$-$4$ at the PBE0/def2-TZVP level are shown in Figure \ref{fig:structures}(a). Other structures can be found in the Supporting Information. From $q=0$ to $q=+3$, the vast majority of structures is composed of fused rings. Indeed, the global minima for $q=$0$-$2 (\textbf{1}, \textbf{1$^+$}, and \textbf{1$^{2+}$}) are all naphthalenic structures, with fused six-membered rings. The situation changes dramatically for higher charge states. In the case of $q=$3, four isomeric structures have enthalpies within 3.3 kcal mol$^{-1}$ from the computed global minimum \textbf{1$^{3+}$}, which contains two terminal C$_3$H$_2$ rings attached to an open-chain C$_4$H$_4$ motif. This structure is also attributed to the global minimum of the tetracationic \ce{C10H8} system, with the most stable structure containing fused rings being 46.0 kcal mol$^{-1}$ higher in energy than \textbf{1$^{4+}$} (\textit{vide infra} for more details).

The second most stable neutral isomer with \ce{C10H8} stoichiometry (\textbf{2}) is the 1-methyleneindene species, which lies 23.4 kcal mol$^{-1}$ above \textbf{1}. This structure is characterised by fused six- and five-membered rings, with a methylene (CH$_2$) group attached to the $\alpha$-carbon of the five-membered ring. This structure is 21.4 kcal mol$^{-1}$ more stable than the 2-methyleneindene isomer, where the CH$_2$ group is attached to the $\beta$-carbon atom of the five-membered ring. The higher stabilisation of the 1-methyleneindene isomer \textbf{2} can be attributed to the formation of a Clar's aromatic sextet, as shown in Figure \ref{fig:structures}(b). Indeed, calculations based on the NBO approach reveal that the difference between the Lewis energy ($E_L$) of the two isomers favours the neutral 2-methyleneindene \textbf{4} with respect to \textbf{2}, as $E_L$(\textbf{2}) $-$ $E_L$(\textbf{4}) $= +$22.3 kcal mol$^{-1}$. The $E_L$ term considers only the Lewis-type NBOs and neglects electronic delocalisation. The correct stability trend between \textbf{2} and \textbf{4} is only obtained if non-Lewis, delocalised contributions ($E_{NL}$) are also taken into account. In this case, $E_{NL}$(\textbf{2}) $-$ $E_{NL}$(\textbf{4}) $= -$44.1 kcal mol$^{-1}$. The sum of the $E_L$ and $E_{NL}$ contributions leads to the full energy difference ($E_{full}$) between the two isomers, which is $E_{full}$(\textbf{2}) $-$ $E_{full}$(\textbf{4}) $= -$21.8 kcal mol$^{-1}$. Following ionisation processes, one could expect that these effects would contribute less to the stability of the systems, decreasing the energy difference between the methyleneindene isomers and eventually inverting the energetic trend. In accordance to such expectations, the singly charged 1-methyleneindene species \textbf{3$^{+}$} is merely 0.9 kcal mol$^{-1}$ more stable than the cationic 2-methyleneindene system \textbf{4$^{+}$}, while the latter structure is more stable than the former for charge states $q=$2$-$4. The dicationic, tricationic, and tetracationic 2-methyleneindene isomers \textbf{2$^{2+}$}, \textbf{2$^{3+}$}, and \textbf{7$^{4+}$} are, respectively, 12.9 kcal mol$^{-1}$, 9.1 kcal mol$^{-1}$, and 6.9 kcal mol$^{-1}$ more stable than the doubly (\textbf{5$^{2+}$}), triply (\textbf{5$^{3+}$}), and quadruply charged (\textbf{9$^{4+}$}, see the Supporting Information) 1-methyleneindene species.

Another important C$_{10}$ backbone is that of the azulene species (\textbf{3}), the most stable neutral \ce{C10H8} system with a seven-membered ring. This structure lies 34.3 kcal mol$^{-1}$ above \textbf{1} and is composed of fused seven- and five-membered rings. The energy difference between the naphthalene skeleton and that of azulene decreases as the systems become more ionised. For $q=1$, the singly charged azulene species (\textbf{2$^{+}$}) is 18.5 kcal mol$^{-1}$ higher than \textbf{1$^{+}$}, an energy drop of more than 50\%. This value decreases to 9.2 and 6.1 kcal mol$^{-1}$ for $q=2$ and $q=3$, respectively. For $q=4$, both structures are more than 50 kcal mol$^{-1}$ above the tetracationic global minimum \textbf{1$^{4+}$}, with the azulene-like structure lying above the naphthalene backbone by around 6 kcal mol$^{-1}$.  

During our investigation, we became interested if structures bearing a pentagonal-pyramidal geometry, such as that of the global minimum of the benene dication, could also be found amongst low-lying \ce{C10H8} isomers. All obtained structures are high energy isomers (see Figure \ref{fig:structures}(c)), indicating that, when the benzene dication global minimum motif is present, the energy always increases.

Amongst the most stable \ce{C10H8} structures, we also found species featuring four-membered rings. Some examples are the neutral dimethylenebenzocyclobutene isomer \textbf{6}, which lies 56.0 kcal mol$^{-1}$ above \textbf{1}, and the tetracationic structure \textbf{5$^{4+}$}, which is composed of non-planar four- and six-membered rings connected by a carbon$-$carbon bond of 1.461 \AA. Molecular structures featuring the aromatic, three-membered cyclopropenyl moiety are significantly more common, particularly for dicationic, tricationic, and tetracationic structures. For $q=2$, the most stable structure featuring the cyclopropenyl ring is \textbf{4$^{2+}$}, which lies 13.9 kcal mol$^{-1}$ above \textbf{1$^{2+}$} and presents such motif connected to a planar \ce{C7H6} seven-membered ring with overall $C_{2v}$ symmetry. For $q=3$ and $q=4$, cyclopropenyl rings are found for species \textbf{1$^{3+}$}, \textbf{3$^{3+}$} (2.7 kcal mol$^{-1}$ above \textbf{1$^{3+}$}), \textbf{1$^{4+}$}, \textbf{2$^{4+}$}, \textbf{3$^{4+}$}, and \textbf{4$^{4+}$} (3.1, 25.0, and 31.0 kcal mol$^{-1}$ above \textbf{1$^{4+}$}, respectively). 

We note that singlet structures featuring eight- to ten-membered rings are not found amongst the low-energy isomers. These include the monocyclic dicationic structures proposed by Leach and coworkers \citep{Leach1989a,Leach1989b} as intermediate species for two-body charge separation of doubly-charged naphthalene and azulene. Selected optimised dicationic structures featuring large rings and their corresponding energies are shown in Figure \ref{fig:rings}. These structures lie from 50 to 80 kcal mol$^{-1}$ above \textbf{1$^{2+}$}. For other charge states, similar structures are also found, but they have even higher energies. Isomer \textbf{14$^{2+}$} is characterised by a quasi-planar structure featuring fused eight- and four-membered rings. Another eight-membered ring structure characterised herein is \textbf{19$^{2+}$}, where two cyclopropene rings are attached to a planar \ce{C8H4} motif forming a $C_{2h}$-symmetry geometry. In turn, \textbf{15$^{2+}$} is composed of an asymmetrical nine-membered ring fused to a cyclopropenyl ring. The most stable ten-membered ring amongst those found in our investigation is \textbf{16$^{2+}$}, where the two hydrogen-free carbon atoms are connected through a formal triple bond of 1.240 \AA. This structure lies 77.0 kcal mol$^{-1}$ above \textbf{1$^{2+}$} and is nearly degenerate to \textbf{13$^{2+}$}, where the hydrogen-free carbon atoms are in \textit{para}-position. Small energy differences are also observed between these structures and \textbf{18$^{2+}$}, which contains the hydrogen-free carbon atoms in \textit{meta}-position. These results indicate that the dicationic ten-membered ring species with \ce{C10H8} stoichiometry are very fluxional. Despite several attempts, no singlet ten-membered ring formally produced from disruption of the central carbon$-$carbon bond of the naphthalene dication \textbf{1$^{2+}$} has been found. Its triplet counterpart (\textbf{$^3$20$^{2+}$}), however, is indeed a minimum energy structure, lying 105.7 kcal mol$^{-1}$ above \textbf{1$^{2+}$}.

After discussing structural properties of the global minima, low- and high-lying energy structures of neutral, singly, and multiply charged \ce{C10H8} isomers, we now focus on particular aspects of their energies and bonding situation. Figure \ref{fig:bonding}(a) shows the structures distribution plot (SDP)  \citep{Sanchez2016} of the species investigated herein in the energy window of 0 to 2.0 eV from the corresponding \ce{C10H8}$^{q+}$ ($q=$ 0$-$4) global minima (1 eV = 23.06 kcal mol$^{-1}$). These energies are significantly smaller than those of the first (S$_0 \rightarrow$S$_1$) electronic transition (4.0 eV) and the ionization potential (8.1 eV) of naphthalene \citep{Cockett1993}. In the SDP plot, each vertical line represents the relative energy position of a stable isomer, with the energies referenced to the global minimum energy structure of each charge state. Amongst the distinct charge states, the population of isomers within the 2.0 eV energy range is the smallest for $q=$0. The neutral charge state also shows the largest energy difference between the global minimum (\textbf{1}) and its first low-lying energy isomer (\textbf{2}). These features highlight the strong stabilising effects at play in the neutral isomers featuring aromatic rings, which are particularly large for naphthalene. 

\begin{figure*}
\centering
\includegraphics[width=\textwidth]{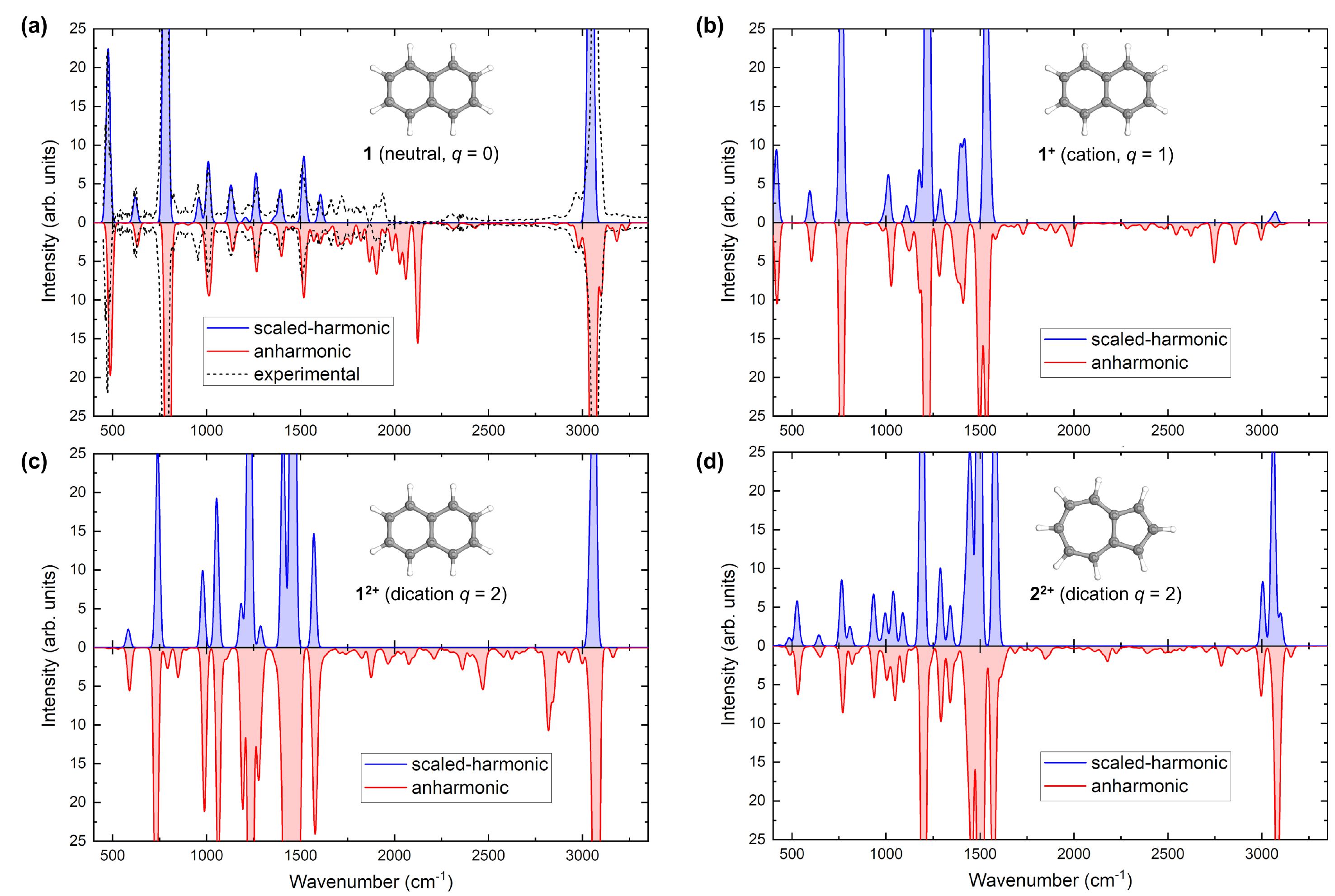} 
    \caption{(a) Comparison of the scaled-harmonic (blue) and anharmonic (red) IR spectra of neutral naphthalene (\textbf{1}). (b), (c) and (d) are the same as (a), but for (\textbf{1$^{+}$}), (\textbf{1$^{2+}$}) and (\textbf{2$^{2+}$}), respectively. The experimental spectrum of \textbf{1} from NIST is also shown with the black dashed line in panel (a). It was obtained in the gas-phase at 300 K.}
    \label{fig:calc_spectra_control}
\end{figure*}

The population of isomers within the 2.0 eV energy window increases for ionised states, reaching its maximum for $q=3$. Isomerisation reactions at the triply charged potential energy surface should, therefore, occur more easily, and the minimum energy structures are expected to have more fluxional character. 

Increasing the charge state to $q=4$ leads to the disappearance of most of the lines and only one low-lying isomer (\textbf{2$^{4+}$}) is found within the 1.0 eV energy range from the tetracationic global minimum \textbf{1$^{4+}$}. Inspection of Figure \ref{fig:bonding}(b) evidences that the structures featuring fused rings with five, six, and seven members, which accounts for the global minimum and low-lying isomers for $q=$ 0$-$3, are particularly unstable for $q=4$ as a consequence of charge repulsion. In fact, even the tetracationic naphthalene \textbf{8$^{4+}$} (see Figure \ref{fig:bonding}, panel c) being a H\"{u}ckel aromatic species, as evidenced by its larger HOMO-LUMO (HL-gap, 2.91 eV) and singlet-triplet (ST-gap, 26.2 kcal mol$^{-1}$) gaps in comparison to those of the antiaromatic \textbf{1$^{2+}$} (2.38 eV and 16.8 kcal mol$^{-1}$, respectively), as well as by its negative NICS(0) ($-5.7$ ppm) and NICS$_{zz}$(0) ($-11.8$ ppm) values, it lies energetically above \textbf{1$^{4+}$} by more than 50 kcal mol$^{-1}$. On the other hand, structures featuring aromatic cyclopropenyl rings and open-chain CH motifs, such as those of \textbf{1$^{4+}$} (see Figure \ref{fig:bonding}, panel d for the corresponding IBOs) and \textbf{2$^{4+}$}, are able to distribute the charge more effectively through the carbon backbone, which explains their particular relative stability for higher values of $q$.

\subsection{Infrared spectra}

In panel (a) of Figure \ref{fig:calc_spectra_control}, we present the scaled-harmonic, anharmonic, and experimental spectra of neutral naphthalene (\textbf{1}), the latter taken from the NIST database \citep{NIST2018}. The calculated peak positions and relative intensities are listed in Table \ref{tab:Bands}. The scaling factors derived for this species are 0.956 for the first group (i.e., $\nu>2500$ cm$^{-1}$), 0.978 for the second group (i.e., $2500>\nu>1111.\bar{1}$ cm$^{-1}$), and 0.973 for the third group (i.e., $\nu < 1111.\bar{1}$ cm$^{-1}$). Overall, the fundamental bands of the harmonic spectrum after scaling are fairly close to the experiment, yielding average errors of only 5.8 cm$^{-1}$ with a maximum value of 18 cm$^{-1}$ associated to one of the C$-$H stretching bands. The harmonic spectrum before scaling shows an error trend that leads to more blueshifted lines as a function of their frequency. However, in relative terms, the blueshift is higher for the bands in the first group, second highest for the bands in the third group, and the least intense for the bands in the second group. This discrepancy pattern is satisfactorily mitigated by the scaling factors. Nonetheless, relatively higher errors associated with the C$-$H stretching bands persist. Considering only the frequency range of groups 2 and 3, the average and maximum errors are reduced to 2.3 and 4.3 cm$^{-1}$, respectively.

\begin{table*}
\begin{center}
\caption{Calculated band positions $\nu_\text{vib}$ (cm$^{-1}$) and relative intensities I$_\text{rel}$ of the fundamental modes of naphthalene (\textbf{1}), (\textbf{1$^{+}$}) and (\textbf{1$^{2+}$}). We compare the calculations to the experimental measurements and give the respective discrepancies $\delta\nu_{\text{vib}}$ (cm$^{-1}$). $ ^{a}$\citep{Mackie2015}, $ ^{b}$\citep{Chakraborty2016}.}
\begin{tabular}{cccccccc}
\toprule
Experimental    &   \multicolumn{3}{c}{Harmonic calculations}                               &   \multicolumn{3}{c}{Anharmonic calculations}                             &   Assignment / symmetry\\
                &   $\nu_\text{vib}$    &   $\delta\nu_{\text{vib}}$    &   I$_\text{rel}$  &   $\nu_\text{vib}$    &   $\delta\nu_{\text{vib}}$    &   I$_\text{rel}$  &\\      
\hline
\multicolumn{8}{c}{\textbf{1} (neutral, $q=0$)}\\
\hline
474$^{a}$       &   476.543             &   2.543                       &   0.199           &   488.462             &   14.462                      &   0.188           &   H$-$C$-$C o-o-p bend i-ph / b$_{3u}$\\
620$^{a}$       &   620.123             &   0.123                       &   0.029           &   631.343             &   11.343                      &   0.030           &   C$-$C str. + H$-$C$-$C i-p bend / b$_{2u}$\\
782$^{a}$       &   782.321             &   0.321                       &   1.000           &   793.512             &   11.512                      &   1.000           &   H$-$C$-$C o-o-p bend i-ph / b$_{3u}$\\
957$^{b}$       &   958.852             &   1.852                       &   0.028           &   1002.883            &   45.883                      &   0.059           &   H$-$C$-$C o-o-p bend o-o-ph / b$_{3u}$\\
1011$^{b}$      &   1009.115            &   1.885                       &   0.070           &   1019.943            &   8.943                       &   0.066           &   C$-$C str. + H$-$C$-$C i-p bend / b$_{2u}$\\
1129$^{b}$      &   1128.541            &   0.459                       &   0.041           &   1138.311            &   9.311                       &   0.028           &   H$-$C$-$C i-p bend / b$_{1u}$\\
1267$^{b}$      &   1262.735            &   4.265                       &   0.057           &   1265.503            &   1.497                       &   0.057           &   H$-$C$-$C i-p bend / b$_{1u}$\\
1389$^{b}$      &   1393.221            &   4.221                       &   0.038           &   1398.787            &   9.787                       &   0.032           &   H$-$C$-$C i-p bend / b$_{1u}$\\
1514$^{b}$      &   1517.102            &   3.102                       &   0.076           &   1515.407            &   1.407                       &   0.069           &   C$-$C str. + H$-$C$-$C i-p bend / b$_{2u}$\\
1603$^{b}$      &   1605.984            &   2.984                       &   0.032           &   1605.447            &   2.447                       &   0.024           &   C$-$C str. + H$-$C$-$C i-p bend / b$_{1u}$\\
3006$^{b}$      &   3023.277            &   17.277                      &   0.049           &   3013.342            &   7.342                       &   0.022           &   Aromatic C$-$H str. / b$_{1u}$\\
3032$^{b}$      &   3025.488            &   6.512                       &   0.007           &   3031.971            &   0.029                       &   0.057           &   Aromatic C$-$H str. / b$_{2u}$\\
3058$^{b}$      &   3040.001            &   17.999                      &   0.467           &   3053.25             &   4.750                       &   0.582           &   Aromatic C$-$H str. / b$_{1u}$\\
3068$^{b}$      &   3050.965            &   17.035                      &   0.347           &   3058.197            &   9.803                       &   0.289           &   Aromatic C$-$H str. / b$_{2u}$\\
\toprule
\multicolumn{8}{c}{\textbf{1$^+$} (cation, $q=1$)}\\
\hline
410                 &   416.754             &   6.754                       &   0.210           &   419.938             &   9.938                       &   0.229           &   H$-$C$-$C o-o-p bend i-ph / b$_{3u}$\\
589                 &   594.893             &   5.893                       &   0.091           &   604.196             &   15.196                      &   0.109           &   C$-$C str. + H$-$C$-$C i-p bend / b$_{2u}$\\
759                 &   763.377             &   4.377                       &   1.000           &   763.826             &   4.826                       &   1.000           &   H$-$C$-$C o-o-p bend i-ph / b$_{3u}$\\
1019                &   1012.563            &   6.437                       &   0.135           &   1027.797            &   8.797                       &   0.179           &   C$-$C str. + H$-$C$-$C i-p bend / b$_{2u}$\\
1121                &   1109.893            &   11.107                      &   0.048           &   1109.852            &   11.148                      &   0.047           &   H$-$C$-$C i-p bend / b$_{1u}$\\
1168                &   1176.030            &   8.030                       &   0.152           &   1178.361            &   10.361                      &   0.183           &   H$-$C$-$C i-p bend / b$_{2u}$\\
1215                &   1220.391            &   5.391                       &   2.228           &   1215.891            &   0.891                       &   2.181           &   C$-$C str. + H$-$C$-$C i-p bend / b$_{2u}$\\
1284                &   1289.241            &   5.241                       &   0.096           &   1286.409            &   2.409                       &   0.101           &   H$-$C$-$C i-p bend / b$_{1u}$\\
1523                &   1530.011            &   7.011                       &   0.869           &   1524.492            &   1.492                       &   0.109           &   C$-$C str. + H$-$C$-$C i-p bend / b$_{1u}$\\
1539                &   1548.630            &   9.630                       &   0.271           &   1535.244            &   3.756                       &   0.170           &   C$-$C str. + H$-$C$-$C i-p bend / b$_{2u}$\\
\toprule
\multicolumn{8}{c}{\textbf{1$^{2+}$} (dication, $q=2$)}\\
\hline
726                 &   740.135             &   14.135                      &   0.348           &   731.427             &   5.427                       &   0.938           &   H$-$C$-$C o-o-p bend i-ph / b$_{3u}$\\
992                 &   979.396             &   12.604                      &   0.132           &   988.371             &   3.629                       &   0.485           &   C$-$C str. + H$-$C$-$C i-p bend / b$_{2u}$\\
1060                &   1053.848            &   6.152                       &   0.256           &   1062.148            &   2.148                       &   0.673           &   H$-$C$-$C i-p bend / b$_{1u}$\\
1184                &   1183.872            &   0.128                       &   0.075           &   1192.328            &   8.328                       &   0.462           &   H$-$C$-$C i-p bend / b$_{2u}$\\
1229                &   1227.573            &   1.427                       &   0.899           &   1234.300            &   5.300                       &   2.200           &   H$-$C$-$C i-p bend / b$_{2u}$\\
1292                &   1287.015            &   4.985                       &   0.037           &   1291.553            &   0.447                       &   0.082           &   H$-$C$-$C i-p bend / b$_{1u}$\\
1409                &   1407.175            &   1.825                       &   0.391           &   1417.240            &   8.240                       &   0.784           &   H$-$C$-$C i-p bend / b$_{1u}$\\
1453                &   1451.843            &   1.157                       &   0.509           &   1449.829            &   3.171                       &   1.514           &   C$-$C str. + H$-$C$-$C i-p bend / b$_{1u}$\\
1461                &   1463.351            &   2.351                       &   1.000           &   1461.650            &   0.650                       &   1.000           &   C$-$C str. + H$-$C$-$C i-p bend / b$_{2u}$\\
1567                &   1571.056            &   4.056                       &   0.195           &   1569.174            &   2.174                       &   0.217           &   C$-$C str. + H$-$C$-$C i-p bend / b$_{2u}$\\
\bottomrule
\multicolumn{8}{l}{str. = stretch; o-o-p = out of plane; i-p = in plane; i-ph = in phase; o-o-ph = out of phase.}
\end{tabular}
\label{tab:Bands}
\end{center}
\end{table*}

Likewise, the relative intensities of the fundamental bands are also in good agreement with laboratory measurements. The most intense vibrational mode in the scaled spectrum is the C$-$H wagging mode at $\sim$782 cm$^{-1}$ (12.8 $\mu$m), followed by the C$-$H stretching modes at $\sim$3055 cm$^{-1}$ (3.3 $\mu$m). The third most intense fundamental band corresponds to a C$-$H bending mode, at $\sim$476 cm$^{-1}$ (21.0 $\mu$m). In order to quantify the agreement between relative intensities, we normalised the theoretical and experimental intensities to their strongest band. The harmonic spectra yielded an excellent average relative intensity ratio of I$_\text{harmonic}$/I$_\text{experimental}\sim0.96$. The relative intensities of the fundamental bands in the anharmonic spectrum are also in very good agreement with the experiment, with I$_\text{anharmonic}$/I$_\text{experimental}\sim0.92$. Its frequencies, however, are not as accurate as the scaled-harmonic counterpart. Considering the whole spectrum, the average and maximum errors were of 9.96 and 45.9 cm$^{-1}$, respectively, which are only marginally adequate to aid astronomical observations. The fundamental modes with b$_{3u}$ symmetry were the least accurately predicted by the anharmonic spectrum, with an average absolute error of $\sim$24 cm$^{-1}$---mainly due to the out-of-plane bending mode at $\sim$1003 cm$^{-1}$ (10.0 $\mu$m), which disagrees with the experimental frequency by almost 46 cm$^{-1}$. Comparatively, the anharmonic b$_{1u}$ and b$_{2u}$ modes are significantly closer to the experiments, with an average error of $\sim$6 cm$^{-1}$. 

The scaled-harmonic and anharmonic spectra of the naphthalene cation (\textbf{1$^{+}$}) are presented in Figure \ref{fig:calc_spectra_control}(b). On panels (c) and (d) of Figure \ref{fig:calc_spectra_control}, we show the spectra of the two least energetic structures of the naphthalene dication (\textbf{1$^{2+}$} and \textbf{2$^{2+}$}). Given the lack of experimental C$-$H stretching bands for the ions, we were only able to derive their scale factors for groups 2 and 3. For the naphthalene cation, the calculated scaling factors are 0.985 (2) and 0.973 (3), whereas for the dication they are both 0.976. The harmonic bands pertaining to the first group were scaled using the scaling factor derived for the neutral species (i.e. 0.956).

\begin{figure*}
\centering
    \includegraphics[width=\textwidth]{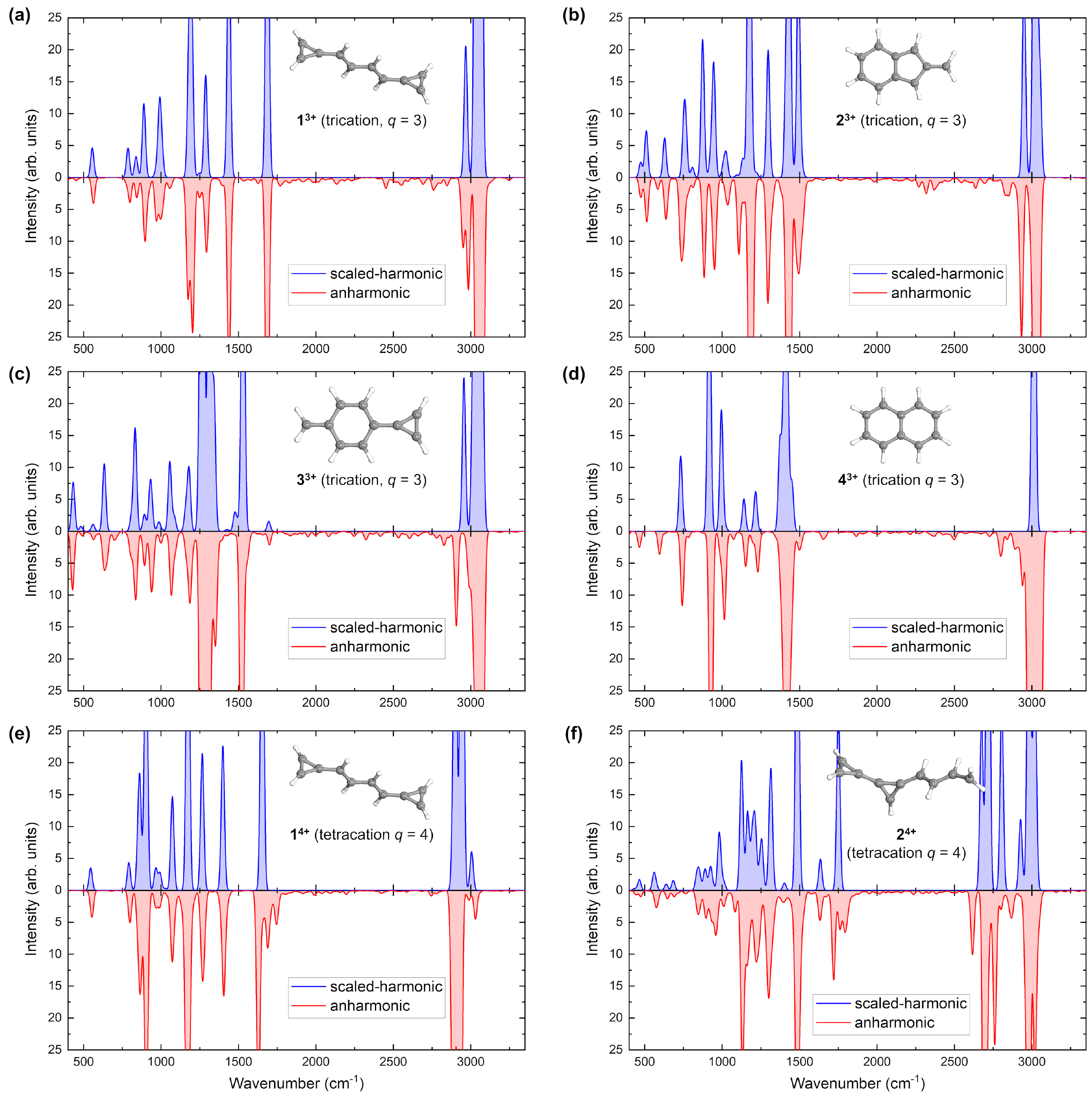} 
        \caption{Scaled-harmonic (blue) and anharmonic (red) spectra of all the remaining calculated naphthalene ions.}
    \label{fig:spectra_rest}
\end{figure*}

The calculated peak positions and relative intensities of the naphthalene cation (\textbf{1$^{+}$}) are listed in Table \ref{tab:Bands}. Differently from the neutral species, the IR spectra of the cation is absent of intense C$-$H stretching bands, in accordance with other experimental and theoretical works \citep{Szczepanski1992,Pauzat1992,Hudgins1994,Langhoff1996,Piest1999}. In the scaled harmonic approximation, its most intense bands correspond to, in decreasing order, a skeletal bending mode at $\sim$1220 cm$^{-1}$ (8.2 $\mu$m), a C$-$H bending mode at $\sim$763 cm$^{-1}$ (13.1 $\mu$m), and a convolution of two skeletal bendings at $\sim$1540 cm$^{-1}$ (6.5 $\mu$m). This spectral profile agrees well with jet-cooled and matrix-isolation experiments \citep{Piest1999}. The scaled frequencies were also considerably close to the experimental spectrum, yielding an average absolute error of $\sim$7 cm$^{-1}$ with a maximum of $\sim$11 cm$^{-1}$. Comparatively, the anharmonic frequency predictions of the fundamental bands were only slightly worse than the scaling approach, yielding average and maximum discrepancies of 6.88 cm$^{-1}$ and 15.2 cm$^{-1}$, respectively. All three IR-active symmetries  (b$_{1u}$, b$_{2u}$ and b$_{3u}$) yield similar errors, i.e., all symmetries are predicted by the anharmonic calculations to a comparable level of accuracy. The resulting spectral profile was very similar to the harmonic approximation, with the two most intense bands at $\sim$1215 cm$^{-1}$ (8.2 $\mu$m) and $\sim$763 cm$^{-1}$ (13.1 $\mu$m) corresponding to the aforementioned skeletal distortion and C$-$H bending modes. Moreover, this spectrum also shows an intense band at $\sim$1500 cm$^{-1}$ (6.7 $\mu$m). However, in contrast with the strong fundamental band that gives rise to this peak in the harmonic approximation, in the anharmonic spectrum this band is mainly composed of two partially-resolved combination bands of relatively high intensity---characterised by the $\nu_{37}$ + $\nu_{35}$ and $\nu_{39}$ + $\nu_{32}$ modes. Moreover, both C$_{10}$H$_8^+$ spectra are overall significantly more intense than those of the neutral species, due to the latter's lack of a permanent electric dipole moment.

The dicationic naphthalene (\textbf{1$^{2+}$}) IR spectra have a slightly more complicated profile than the singly charged counterparts. Its calculated peak positions and relative intensities are listed in Table \ref{tab:Bands}. Both the scaled-harmonic and the anharmonic spectra show an intense double band at $\sim$1500 cm$^{-1}$ (6.7 $\mu$m) due to out-of-phase H$-$C$-$C bending and C$-$C stretching modes, as well as another intense band at $\sim$1250 cm$^{-1}$ (8.0 $\mu$m) related to in-phase H$-$C$-$C bending. They also show moderately intense bands at $\sim$730 cm$^{-1}$ (13.7 $\mu$m), $\sim$1400 cm$^{-1}$ (7.1 $\mu$m), and $\sim$3100 cm$^{-1}$ (3.2 $\mu$m), respectively related to wagging, rocking, and C$-$H stretching modes. Comparatively, the naphthalene cation rocking bands are significantly less intense. Moreover, combination bands of moderate intensity in the $\sim$1400--1600 cm$^{-1}$ (6.3--7.1 $\mu$m) range considerably change the profile of the spectrum in comparison to the harmonic approximation. In fact, anharmonic corrections seem to play an important role in the IR spectra of doubly charged naphthalene (\textbf{1$^{2+}$}) \citep{Banhatti2021}: the average and maximum frequency errors of the anharmonic spectra were, respectively, 3.9 and 8.3 cm$^{-1}$--- $\sim$20\% and $\sim$41\% less than the ones of the scaled-harmonic spectrum. These errors are consistent throughout the entire frequency range of groups 2 and 3 (2500 cm$^{-1}>\nu$), and are independent of the mode symmetry. Both scaled-harmonic and anharmonic spectra are reasonably to significantly close to the experimental spectrum of \cite{Banhatti2021} at the measured frequency range.

The scaled-harmonic and anharmonic spectra of the remaining calculated structures with $H^{298}<$10 kcal mol$^{-1}$ are shown in Figure \ref{fig:spectra_rest}. Due to the lack of experimental data, the harmonic bands of all ionic species with charges above $q$ = +2 were scaled using the factors derived for the naphthalene dication \textbf{1$^{2+}$}. The profiles of the spectra change appreciably with the charge state, even for similar naphthalene-like geometries. Different structures of the same charge states also result in very distinct spectra. Combination bands of intermediate to high intensity were derived in the majority of the anharmonic spectra, but the resulting spectral profiles remained mostly unchanged---or only slightly different---due to decreases in the intensity of fundamental bands that fall in their vicinity.

As observed for the dication global minimum (\textbf{1$^{2+}$}), the IR spectra of the remaining multiply charged species are considerably more complex than those of the neutral and singly charged species, which hampers the characterisation of potential fingerprint regions. With the exception of the naphthalene cation, all charged species show C$-$H stretching features of moderate to high strength. In fact, for most of the trication structures and all of the tetracation structures, these are the dominant bands of the spectra. The tetracationic, second-lowest-energy \ce{C10H8} structure \textbf{2$^{4+}$} is unique in that it has an aliphatic extremity that results in C$-$H stretching bands with frequencies ranging from 2800 to 3000 cm$^{-1}$ (3.3--3.6 $\mu$m) additionally to the higher-energy features typical of unsaturated species. The \textbf{3$^{3+}$} (third-lowest-energy structure) of the trication species also has aliphatic C$-$H bonds, but the resulting IR features have higher frequencies at around 3090 cm$^{-1}$ (3.2 $\mu$m). This shift is likely due to the close proximity to the aromatic ring.

For the neutral species, the out-of-plane C$-$H modes that fall in the frequency range of $\sim$666--1000 cm$^{-1}$ (10--15 $\mu$m window) are the dominant bands in the spectrum. Comparatively, the in-plane C$-$H bending and C$-$C stretching modes that fall in the frequency range of 1000--2000 cm$^{-1}$ (5--10 $\mu$m window) are not as important in terms of intensity. However, this profile changes for the cation and dication species: the latter modes become the dominant ones in the spectra. Based on this well-known phenomenon, the intensity ratios of the bands in the 5--10 $\mu$m and 10--15 $\mu$m regions have been widely used as both qualitative and quantitative proxies to track the charge balance of PAH populations in astronomical sources \citep{draine&li01, Boersma2015, Boersma2016, Maragkoudakis2020}. Indeed, this trend has been observed for the most part of the cationic species of Figures \ref{fig:calc_spectra_control} and \ref{fig:spectra_rest}. In spite of that, the IR spectra of the trication naphthalene-like structure (\textbf{4$^{3+}$}) (see d panel of Figure \ref{fig:spectra_rest}) do not reproduce this behaviour. In fact, its out-of-plane C$-$H modes are more intense than the in-plane C$-$H bending and C$-$C stretching modes, which better resembles the spectral profile of the neutral species.

\section{Astrophysical implications}
\label{implications}

\begin{figure*}
\centering
\includegraphics[width=0.9\textwidth]{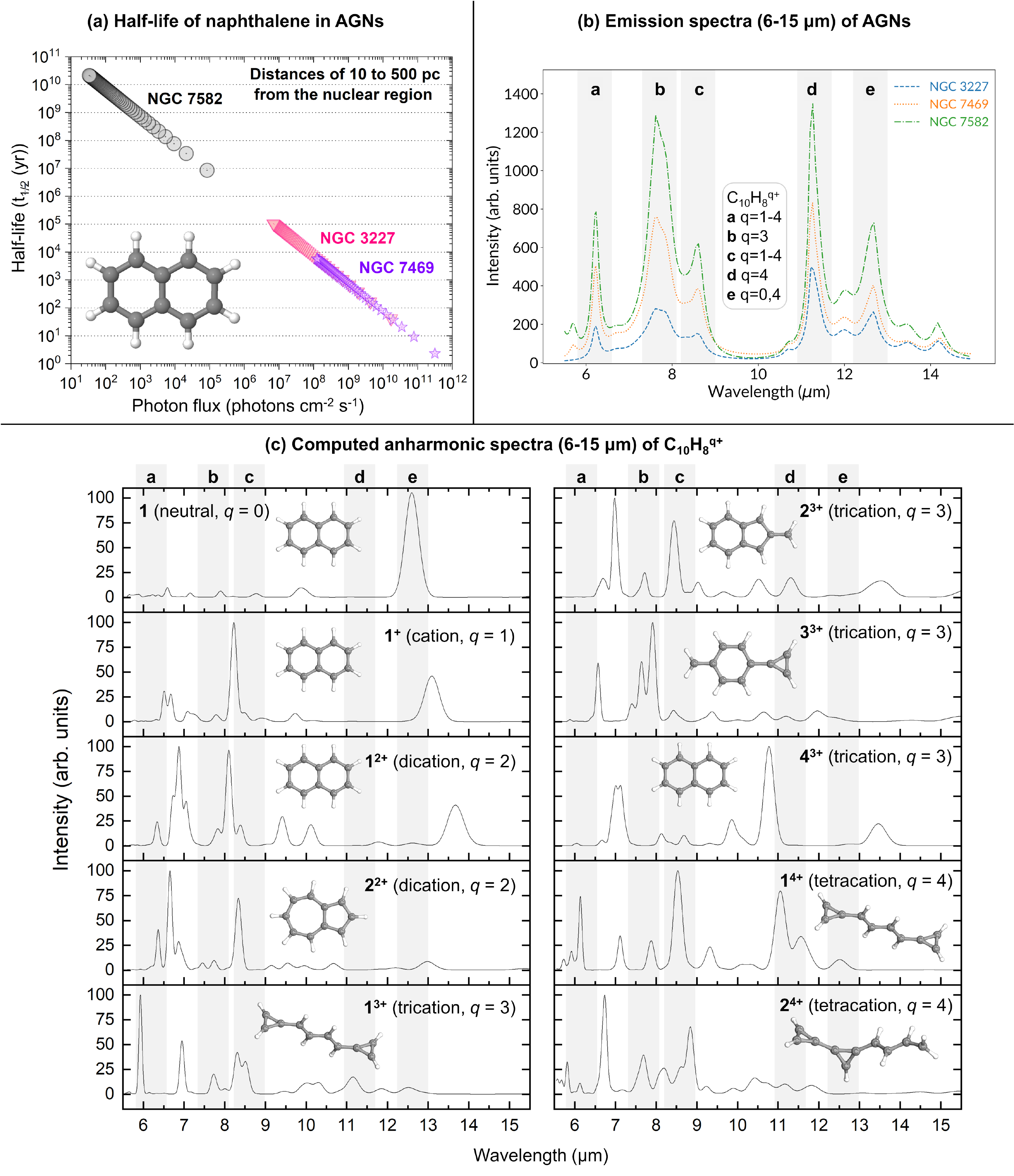}  
    \caption{(a) Half-life values of the PAH molecules studied herein after interaction with the X-ray radiation field of selected AGNs. Each data point refers to a specific distance between 10–500 pc from the Seyfert nucleus. (b) Mid-IR IRS/Spitzer spectra of the Seyfert galaxies that were analysed in this work, where the shaded areas represent the wavelength regions of the main PAH features. (c) Spectra of singly, doubly, and thirdly charged naphthalene scaled and compared to selected AGNs.}
    \label{fig:half-life}
\end{figure*}

In this section, we discuss the present results in the context of the chemistry of circumnuclear regions of AGNs. From the observed X-ray luminosities, we estimate the half-lives of naphthalene in the circumnuclear regions of the selected AGN sources, whose results are shown in Figure \ref{fig:half-life}(a). As discussed by \cite{Monfredini2019}, multiply charged naphthalene molecules are indeed present in the mass spectrum of the system after interaction with 2.5 keV photons, revealing that such species could contribute to the emission spectrum of objects featuring high energetic fluxes, such as AGNs. The mid-IR emission spectra of the selected AGN sources, namely NGC 3227, NGC 7469 and NGC 7582, are shown in Figure \ref{fig:half-life}(b). They were compiled by the ATLAS project \citep{hernan&hatziminaoglou11} and are also available through IRSA (NASA/IPAC Infrared Science Archive; https://irsa.ipac.caltech.edu/). The main spectral features are compared with the anharmonic vibrational transitions of the most representative C$_{10}$H$_8^{q+}$ species calculated herein, whose computed data in the 6-15 $\mu m$ range is shown in Figure \ref{fig:half-life}(c). 

UIR bands - and therefore their carriers, which include PAHs - are widely observed in the universe, even in highly dissociative regions such as the circumnuclear regions of AGNs \citep[e.g.][]{Alonso-Herrero2020}. To explain the emission of PAH around AGN, \cite{Voit1992} proposed that these molecules should be protected by a dense torus surrounding the highly ionising central source. Knowing that the gas entering a central AGN can form a circumnuclear disk, one could expect that this structure would play an important role in obscuring the AGN \citep[][and references therein]{Netzer15}. 
There is in fact evidence of PAH detection in the proximities of type-1 Seyfert and intermediate-type Seyfert galaxies \citep{Sales2010, Jensen2017}, suggesting that PAHs can on the one hand survive in these regions and, on the other, that PAH excitation is not necessarily limited to star-forming regions.

ISM dust grains follow a size distribution in the range of 0.4 nm to larger than 10 nm \citep{tielens08}, with PAHs at the smallest end being characterised in terms of the number of carbon atoms contained \citep{galliano+08}.
Despite occupying the small end of dust size distribution, PAHs are large molecules and their vibrational modes result in broad emission features in the mid-IR region associated to the different stretching and bending modes of C$-$H and C$-$C bonds, mainly at the $\sim$3-13 $\mu$m region \citep{allamandola+89}.
As shown by e.g. \cite{draine&li01} and \cite{li&draine01}, some of these features can be used to trace the structure, size, and charge of the molecules. It has been shown that neutral and singly ionised PAHs are not sufficient to model the features and multiply ionised PAHs must be accounted for in the ISM to provide good matches between laboratory and astrophysical data \citep{bakes+01a}. Based on \cite{bakes+01b}, the features of multiply charged PAHs are more pronounced in the C$-$C modes in comparison to the C$-$H modes. The possibility of identifying ionised PAHs in the ISM has been proposed and discussed for several decades. In this regard, \cite{Leach1986} showed the importance of studying the photostabilities of PAH monocations and dications and their astrophysical implications. Subsequently, \cite{bakes+01a} proposed that the probability of finding PAHs in a state of multiple ionisation increases with their molecular size. In regions with high energy fluxes, PAHs, therefore, might be doubly, triply, or even more highly charged. From the experimental point of view, only some of the simplest ionised species have been extensively studied in the astrophysical context \citep{Szczepanski1993, Ruiterkamp2002, tielens08, Zhen2015, Banhatti2021}. PAHs with higher charge states are significantly more challenging to study both from the experimental and computational points of view, and their unambiguous detection in astrophysical environments is still unachievable \citep{Zhen2017}.

\subsection{Half-lives of naphthalene in PAH-containing AGN sources}

In order to compare our spectroscopic data with astrophysical objects, we selected three distinct PAH-containing AGNs, namely NGC 3227, NGC 7469, and NGC 7582, taken from \cite{Jensen2017}. These AGNs are classified as intermediate Seyfert-type galaxies, with PAH bands identified in mid-IR and predominance of small, neutral, and pure PAH species \citep{MartinsFranco2021}. The X-ray luminosities in the range of 2-10 keV of these selected AGNs are shown in Table \ref{tab:agn}. 

\begin{table}                                           
\centering                                           
\caption{Some properties of the AGN sources studied herein. $L_{X}$ (eV s$^{-1}$) stands for the X-ray luminosities integrated from 2-10 keV. $N_H$ (cm$^{-2}$) is the column density of atomic hydrogen. $\tau_x$ and $\xbar{F}_X$ (photons cm$^{-2}$ s$^{-1}$) are the estimated X-ray optical depths and the average X-ray photon fluxes at 2.5 keV within distances of 10-500 pc from the Seyfert nucleus of the corresponding sources.}  
\label{tab:agn}                                           \begin{tabular}{lccccc}                          
\toprule                                           
Source & Type & {$L_X$}$^a$ &  {$N_H$}$^b$ &  $\tau_x$ & {$\xbar{F}_X$} \\                                    
 &  & (eV s$^{-1}$) &  (cm$^{-2}$) &   & (cm$^{-2}$ s$^{-1}$) \\  \midrule                                  
NGC 3227 & S1.5 & 8.62$\times10^{53}$   & 1.86$\times10^{22}$ & 0.52 & 5.57$\times10^{8}$ \\                                  
NGC 7469 & S1.5 & 9.67$\times10^{54}$    & 1.15$\times10^{21}$ & 0.03 & 1.02$\times10^{10}$ \\          
NGC 7582 & S1i & 2.11$\times10^{54}$ & 4.90$\times10^{23}$ & 13.64 & 2.73$\times10^{3}$   \\            
\bottomrule 
\multicolumn{6}{l}{$^a$Taken from \cite{Jensen2017}}\\
\multicolumn{6}{l}{$^b$Taken from \cite{Jaffarian2020}}\\
\end{tabular}                                           
\end{table}

NGC 3227 is a spiral-type galaxy interacting with a neighboring dwarf elliptical galaxy \citep{Delgado1997}. This object presents a large concentration of molecular gas in the central part, equivalent to 80\% of the total CO emission that is detected in a radius of $\sim$1.2 kpc \citep{Delgado1997}. NGC 3227 is located at a distance of $\sim$22 Mpc, at a luminosity of \textit{log} L$_X$ = 42.14 $\pm$ 0.21 erg s$^{-1}$ \citep {Jensen2017}. The galaxy also exhibits a short-term hard X-ray variability likely related to variation in the intrinsic continuum emission \citep{RamosAlmeida2009}. 

In turn, NGC 7469 is an active core spiral galaxy with a large-scale stellar bar (several kpc) identified in the near-IR \citep{Knapen2000}. Its circumnuclear region is surrounded by a 1".5-2".5 starburst ring \citep{Soifer2003}. In the mid-IR, the most central region of NGC 7469 ($\sim$1" of the nucleus) is dominated by emissions in the K band (2.2 $\mu$m) and the 3.3 $\mu$m band \citep{Izumi2015}, the latter indicating a strong presence of PAHs in a region very close to the nucleus. NGC 7469 is at $\sim$ 67.9 Mpc and has a luminosity of \textit{log} L$_X$ = 43.19 $\pm$ 0.07 erg $s^{-1}$ \citep{Jensen2017}. 

Lastly, NGC 7582 is a nearby spiral galaxy, located at a distance of $\sim$23.0 Mpc and a luminosity of \textit{log} L$_X$ = 42.53 $\pm$ 0.38 erg s$^{-1}$ \citep{Jensen2017}. NGC 7582 presents a circumnuclear region of active star formation and an AGN-powered outflow, making it an ideal site for studies focusing on the feeding and feedback processes in which AGN lead to an increase or decrease in star formation \citep{Riffel2009}.

In order to estimate the survival of naphthalene in the proximity of the circumnuclear regions of the aforementioned AGN sources, we used the experimental data taken from \cite{Monfredini2019} at photon energy E = h$\nu$ = 2.5 keV to obtain the photodissociation cross-section, $\sigma_{ph-d}$ (cm$^{2}$), of the molecule. For that, the average X-ray photon flux,  $\xbar{F}_X$ (photons cm$^{-2}$ s$^{-1}$) of each source, in a fixed range of distances from the circumnuclear region (10-500 pc), was calculated using the following equation:

\begin{equation} \label{eq:flux}
\xbar{F}_X = \frac{L_{X}}{4\pi r^{2}h\nu}e^{-\tau_{x}},
\end{equation}

\noindent where $L_X$ is the X-ray luminosity (eV s$^{-1}$) integrated from 2 to 10 keV, $r$ is the distance from the nucleus to a position between 10 to 500 pc and $\tau_{x}$ is the X-ray optical depth given by:

\begin{equation} \label{eq:tsx}
\tau_{x}= \sigma_{\text{H}}(E) N_{\text{H}}, 
\end{equation}

\noindent where $ \sigma_{H}$(E) and $N_{\text{H}}$ are the X-ray photoabsorption cross section and the column density of the hydrogen atom, respectively.
The values of $\sigma_{H}$(E) can be obtained by \citep{Gorti2004}:

\begin{equation} \label{eq:csx}
\sigma_{\text{H}}(E) = 1.2 \times10^{-22} \left(\frac{E}{1 \text{keV}}\right )^{-2.594}.  
\end{equation}

\noindent Therefore, at a 2.5 keV photon energy, $\sigma_{H}$ = 1.11 $\times 10^{-23}$ cm$^{2}$. 

The calculated $\xbar{F}_X$ values are shown in Table \ref{tab:agn}. Now, taking the X-rays photon flux values and using the photodissociation cross-section values from \cite{Monfredini2019}, we estimate the half-life of naphthalene using the following equation (for more details see \citealt{Monfredini2019}):

\begin{equation}\label{eq:hl}
t_{1/2}=\frac{ln \, 2}{k_{ph-d}}, 
\end{equation}

\noindent where $k_{ph-d}$, the photodissociation rate (s$^{-1}$), is equal to the product of $\sigma_{ph-d}$ and $\xbar{F}_X$ at the same photon energy. The average half-lives of naphthalene for each source vary from 7.4$\times 10^{9}$ yr for NGC 7582 to 2.0$\times 10^{3}$ yr for NGC 7469 (see Figure \ref{fig:half-life}(a)). In the case of NGC 7582, the half-life times obtained show that naphthalene can present comparable times, or even higher, than that necessary for the injection of PAHs in the interstellar medium (2.5$\times 10^{9}$ yr, \citealt{Jones1994}). The conditions of NGC 7582 are optimal for the formation of these molecular species and this is due to the fact that, although its x-ray luminosity of 2.11$\times 10^{54}$ eV s$^{-1}$ is considerably large, its high $N_H$ value of 4.90$\times 10^{23}$ cm$^{-2}$ makes it the most obscured AGN source amongst those investigated herein. Therefore, we can speculate that the thick dust surrounding the circumnuclear region is an effective blocker, which exerts a high degree of protection for PAHs, including smaller PAHs such as naphthalene.

On the other hand, for the other two AGNs, the half-lives obtained are smaller than the PAH injection time. For NGC 7469, the ratio between the PAH injection time and the average half-life of naphthalene is as large as $\sim$10$^{6}$, even though the X-ray luminosity of this source is higher than that of NGC 7582 by a factor of $\sim$ 4.6. This is due to the fact that the column density $N_H$ in NGC 7469 is particularly thin ($\tau_x = 0.03$, see Table \ref{tab:agn}), around 450 times smaller than that of NGC 7582 ($\tau_x = 13.64$). In order to account for the presence of PAHs, particularly those of smaller sizes, in such objects that combine highly dissociative fluxes and optically thin column densities, one or more PAH protection mechanisms (see \citealt{Quitian-Lara2018,Monfredini2019}) should be at play in order to reduce the effectiveness of dissociation induced by photon-molecule interaction. 

\subsection{Broadband emission features in the mid-IR spectra of AGNs}

\begin{figure}
\centering
    \includegraphics[width=\columnwidth]{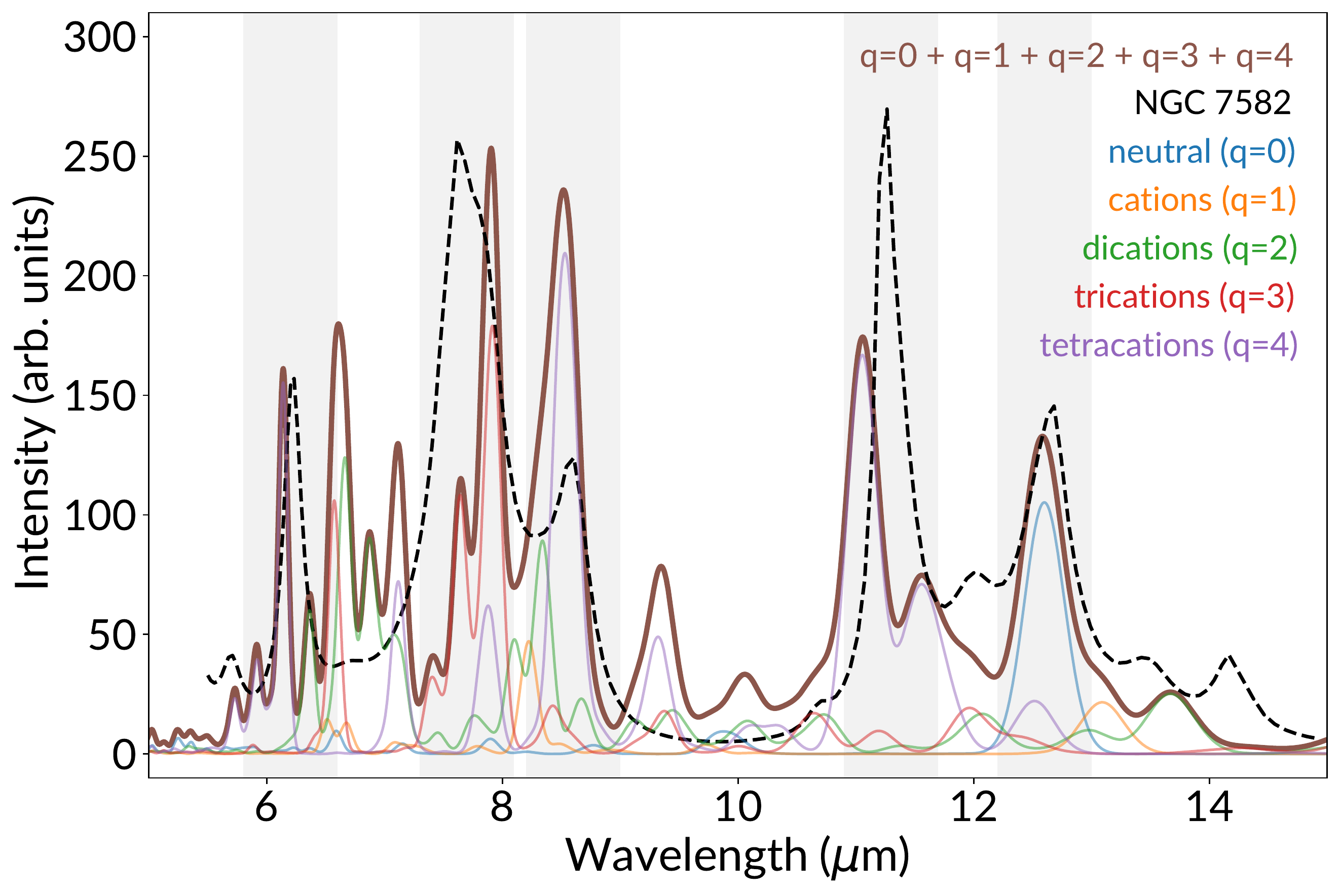} 
    \caption{Comparison of the NGC 7582 mid-IR spectrum with a theoretical curve obtained by a linear combination of the computed anharmonic spectra of distinct C$_{10}$H$_8^{q+}$ species. For a better visualisation, the computed spectra were scaled.}
    \label{fig:AGN}
\end{figure}

In Figure \ref{fig:half-life}(c) we display naphthalene ionisation states addressed in this work with their respective structures and spectra. The first feature (\textbf{a}), appearing at 6.2 $\mu$m, is associated with the C$-$C stretching band of PAHs and is common to the main C$_{10}$H$_{8}^{q+}$ structures with $q=$ 0$-$4. However, as expected due to their higher dipole moments, the intensity contributions coming from the ionised species are larger than that of the neutral counterparts. The second feature (\textbf{b}) coincides with the 7.7$ \mu$m emission band. This feature is broadly considered to be the combination of two emission bands, at 7.6 $\mu$m and 7.8 $\mu$m, the relative intensities of which are employed to classify different sources \citep{Canelo+21}. All AGNs investigated herein are characterised by the larger intensity of the 7.6 $\mu$m component, thus pertaining to the spectral class A' \citep{peeters+02}. This component is indeed consistent with PAH emission, which in this wavelength is originated mostly from in-plane C$-$H bending modes. For the C$_{10}$H$_{8}^{q+}$ species, vibrations within this band are particularly intense for $q=3$, which presents strong features at 7.7 and 8.6 $\mu$m. The latter is also assigned to C$-$H in-plane modes and is seen in feature (\textbf{c}). The features (\textbf{d}) and (\textbf{e}), 11.3 $\mu$m and 12.7 $\mu$m respectively, are associated with C$-$H out-of-plane bending modes and, as predicted, are virtually unseen for ionised C$_{10}$H$_{8}^{q+}$ species, except for $q=4$. In opposition to the other ionised C$_{10}$H$_{8}^{q+}$ species, the tetracation presents a magnified flux at higher wavelengths---and especially at  feature (\textbf{d}). Finally, the C$-$H out-of-plane mode at 12.7 $\mu$m is the most intense vibrational contribution of neutral naphthalene, and coincides with the feature (\textbf{e}) present in all AGN sources investigated herein. Regarding the ionised naphathalenic structures, this band becomes increasingly redshifted as the charge state increases (see also Table \ref{tab:Bands}). Consequently, the peak position of band 12.7 $\mu$m could potentially serve as a tracer of the medium's ionization state. Finally, we compare the NGC 7582 mid-IR spectrum with a composed theoretical curve obtained by a linear combination of the computed anharmonic spectra of Figure \ref{fig:half-life}(c). These results are summarised in Figure \ref{fig:AGN}. As expected, the 12.7 $\mu m$ band is mainly reproduced by the neutral naphthalene. Interestingly, we also obtained that the combination of highly ionised C$_{10}$H$_8$ species is able to reproduce some features of the 6.2 $\mu m$ and 7.7 $\mu m$ bands, with trications contributing mainly to the higher-wavelength portion of the latter and tetracations to the lower-wavelength portion of the former. These findings hint at the importance of considering not only the charge state of PAHs to the composition of AGN IR bands, but also the distinct low-lying isomeric structures that might be produced following multiple ionisation events. However, the small size of our prototypical model is not large enough to provide definite answers. Therefore, further investigations of this phenomenon with larger multiply charged PAH species are warranted.

The mid-IR broadband features have been  explored in the literature as a diagnostic tool for the ISM's radiation field and the ionisation state of PAHs \citep[e.g.][]{lambrides+19, MartinsFranco2021}. \cite{draine&li01} demonstrate that the 11.3-to-7.7 $\mu$m band flux ratios can tells us something about the overall characteristics of the PAH population in the source: the former is associated to the ionisation fraction, whereas the latter is a strong function of the PAH size. This correlation, however, is not observed for the C$_{10}$H$_{8}^{q+}$ species explored here (see Figure S6 in the Supporting Information). In fact, we detect no trend for the IR band intensities with respect to the molecule's charge state nor structure. That is rather unsurprising, considering that the model of \cite{draine&li01} was constructed for what the authors called a "generic PAH" with C$>$16 and that is either neutral or singly ionised. Moreover, their model considers environments that are not completely compatible with AGNs, such as the cold neutral and warm ionised ISM, as well as photodissociation regions. In this sense, further studies that focus on the IR spectra of multiply ionised PAHs of different sizes are warranted to derive adequate proxies for higher ionisation states.

\section{Conclusions}
\label{conclusions}

We investigated the geometric properties and patterns of neutral, singly and multiply charged naphthalene and its C$_{10}$H$_{8}$ isomers by searching extensively for molecular structures with charge states $q$ = 0$-$4. The most stable C$_{10}$H$_{8}$ isomers with $q$ = 0$-$2 are predominantly composed of fused rings. However, for higher charge states, this profile changes sensibly, with open chains dominating the low-energy structures of the tetracationic C$_{10}$H$_{8}$. The effect of electronic delocalisation is shown to become gradually less important to the stabilisation of the systems as the charge state increases, and the increase in charge state also results in a significantly smaller energy difference between naphthalenic and azulenic backbones. High-energy pentagonal-pyramidal structures fused to six- or five-membered rings were found for the C$_{10}$H$_{8}$ systems with $q$ = 2$-$4, whereas for the species with lower charge states they were not observed. Additionally, four- and three-membered rings were also repeatedly observed, with the latter being notably common for the tetracationic systems. However, regarding larger rings with eight- to ten members, we do not find singlet structures with those features amongst the low-energy isomers. 

The neutral charge state is evidently highly stabilised by the presence of aromatic rings, as is conveyed by the isomer population and their energy difference in the SDP plot. For higher charge states, and especially for $q$ = 3, the population of isomers increases significantly. Thus, isomerisation reactions should be more frequent and consequently play a more important role in the characterisation of these systems. For $q$ = 4, charge repulsion leads to the destabilisation of structures with five-, six- and seven-membered fused rings, whereas aromatic cyclopropenyl rings and open-chain CH motifs increase the system’s stability by allowing to distribute the charge more effectively.

We also assessed the spectroscopic properties of the most stable structures ($H^{298.15}<10$ kcal mol$^{-1}$) in the infrared region, with the goal of aiding in the interpretation of astronomical observations. We simulate both their harmonic and the anharmonic IR spectra, and derive scaling factors in accordance to the PAHdb guidelines based on the experimentally-measured bands of the neutral, singly, and doubly charged global minima. Overall, the scaled-harmonic spectra were reasonably close to experimental results (when available), with average errors well within 10 cm$^{-1}$ from the measured values. Comparatively, the anharmonic spectrum of neutral naphthalene was sensibly less accurate than the scaled-harmonic counterpart. For higher charge states, however, anharmonic corrections become increasingly more important, eventually yielding a more accurate spectrum than the scaled-harmonic approach for the dicationic species. Nonetheless, all spectra obtained here were satisfactorily close to experiments and thus both approaches seem adequate to aid astronomical observations.

The computed spectra of the remaining low-energy C$_{10}$H$_{8}$ structures are also provided. Overall, the complexity of the infrared emission spectra increases significantly for higher ionisation states. Intense bands in the C$-$H stretching region are consistent for the vast majority of the species studied here, with the only exception of the naphthalene cation (\textbf{1$^{+}$}). Furthermore, the species with higher charge states also typically present spectra dominated by the bands in the 5$-$10 $\mu$m region. The naphthalenic trication (\textbf{4$^{3+}$}), however, does not follow this trend, resembling more closely the profile of the neutral species.

The hard radiation generated by supermassive black holes accreting material may lead to the formation of singly, doubly, and even multiply charged PAHs, including in the case of C$_{10}$H$_{8}$. Therefore, the species explored in this work are potentially part of the chemistry of circumnuclear regions of AGNs, and could thus be enriching their emission spectra. For this reason, we explore the survival of naphthalene in the proximity of a set of PAH-containing AGNs, using our previous experimental value of photodissociation cross-section, $\sigma_{ph-d}$, and its half-life, $t_{1/2}$, at photon energy of 2.5 keV. For NGC 7582, the naphthalene half-life is comparable to, or even higher than the time necessary for the injection of PAHs in the interstellar medium, which is likely due to the shielding of the PAHs from the high X-ray flux by the dust surrounding the circumnuclear region. For the rest of the selected sources, however, the dust available is not as effective in shielding the PAH molecules; other protection mechanisms, such as PAH superhydrogenation \citep{Reitsma2014,Quitian-Lara2018,Stockett2021} and PAH-to-dust adsorption and incorporation processes \citep{Monfredini2019}, are required.

We also compared the computed IR spectra of the most representative low-energy C$_{10}$H$_{8}$ structures with that of the selected AGNs. The objects' spectra are consistent with PAH emission, and are especially compatible with the IR bands of the $q$ = 3 and $q$ = 4 species. The C$-$H out-of-plane modes of the naphthalenic species present increasingly redshifted frequencies as a function of the charge state, pointing in the direction of a promising tracing capability of the 12.7 $\mu$m band regarding the ionization state. Finally, we assessed the capability of typical ionisation-state proxies to predict the charge of the multiply ionised C$_{10}$H$_{8}$, but found no correlation. Thus, further works dedicated to exploring the IR band ratios of PAHs with multiple charge states are desirable.

\section*{Acknowledgements}

 J.C.S., H.M.B.-R., F.F. and R.R.O. acknowledge the National Laboratory for Scientific Computing (LNCC$/$MCTI, Brazil) for providing HPC resources of the SDumont supercomputer, which have contributed to the research results reported within this paper (http:$//$sdumont.lncc.br). F.F. thanks Julius-Maximilians-Universit\"{a}t W\"{u}rzburg and Prof. Dr. Bernd Engels for providing additional computational resources. H.M.B.-R. and R.R.O. acknowledges FAPERJ and CNPq for financial support.  This study was financed in part by the Coordenação de Aperfeiçoamento Pessoal de Nível Superior (CAPES) - Finance Code 001.

\section*{Data Availability}

Data available on request.

\bibliographystyle{mnras}
\bibliography{ref}

% Don't change these lines
\bsp	% typesetting comment
\label{lastpage}
\end{document}